\documentclass{article}
\usepackage{spconf,amsmath,graphicx,booktabs, xcolor, tabularx}
\usepackage{footmisc}
\usepackage{subfiles}
\usepackage[hyphens]{url}
\usepackage{hyperref}
\usepackage{breakurl}
\usepackage{xcolor}
\definecolor{Highlight}{HTML}{39b54a}  
\definecolor{darkred}{RGB}{139, 0, 0}
\definecolor{darkgreen}{RGB}{0, 100, 0}

\newcommand{\CellWithForceBreak}[2][c]{
\begin{tabular}[#1]{@{}c@{}}#2\end{tabular}}
\usepackage{multirow}
\usepackage{hyperref}

\title{IMPROVING RARE-WORD RECOGNITION OF WHISPER IN ZERO-SHOT SETTINGS}

\name{Yash Jogi$^\dagger$, Vaibhav Aggarwal$^\dagger$, Shabari S Nair, Yash Verma, Aayush Kubba}

\address{Sprinklr, India \\
         {\tt\small \{yash.jogi, vaibhav.aggarwal, shabari.nair, yash.verma, aayush.kubba\}@sprinklr.com}}
\begin{document}

\maketitle
{\let\thefootnote\relax\footnotetext{$^\dagger$Equal contribution.}}

\begin{abstract}
Whisper, despite being trained on $680K$ hours of web-scaled audio data, faces difficulty in recognizing rare words like domain-specific terms, with a solution being contextual biasing through prompting. To improve upon this method, in this paper, we propose a supervised learning strategy to fine-tune Whisper for contextual biasing instruction. We demonstrate that by using only $670$ hours of Common Voice English set for fine-tuning, our model generalizes to $11$ diverse open-source English datasets, achieving a $45.6\%$ improvement in recognition of rare words and $60.8\%$ improvement in recognition of words unseen during fine-tuning over the baseline method. Surprisingly, our model's contextual biasing ability generalizes even to languages unseen during fine-tuning.

\end{abstract}


\section{Introduction}
\label{sec:intro}

Recent years have witnessed a growing interest in training end-to-end speech recognition systems on large-scale audio data, leading to robust and generalized speech recognition models \cite{wav2vec2,whisper}. Among-st such models, Whisper \cite{whisper} stands out as the only open-source speech recognition model trained on a massive scale of $680,000$ hours of web-scraped audio data on various speech related tasks, achieving low Word-Error-Rate (WER) across diverse domains and languages. Hence, such an open-source model serves as a foundational model in the domain of speech and has been used ``out of the box" in a variety of applications \cite{application, application2, application3}. Despite being trained on such a large-scale dataset, Whisper struggles with the recognition of rare words such as proper nouns or domain specific words, which might be sparse or absent from its training data \cite{tcpgen}.

This issue of difficulty in recognizing rare words has been prevalent among other end-to-end Automatic Speech Recognition (ASR) models as well \cite{contextrnnt, personalizationctc}. To address this challenge, contextual biasing has been used extensively as one of the most popular solutions \cite{sathyendra2022contextual, tang2024improving}. Specifically, this approach involves providing relevant contextual knowledge in the form of a list of words or phrases which can be contact names, application names, a list of medical terms, or any other domain specific words to an ASR model, so as to make their recognition more accurate. For example, in the medical domain, an ASR model trained on general data might struggle to transcribe rare terms like ``spirometry''. However,  providing relevant context, such as a list of medical terms, can significantly reduce WER and enhance the model's utility and reliability.

A key difference between Whisper and previous ASR architectures such as wav2vec2.0 \cite{wav2vec2} or Transformer-Transducer \cite{transformertrans} is Whisper's unique prompt functionality. Specifically, Whisper differs from other ASR models in that it enables transcription control through prompting. As suggested in \cite{whisper}, OpenAI's official documentation for Whisper \footnote{\label{myfootnote}\href{https://platform.openai.com/docs/guides/speech-to-text/prompting/}{https://platform.openai.com/docs/guides/speech-to-text/prompting/}} mentions several ways to use the prompt feature, particularly to increase the recognition accuracy for rare words by including such a word list in the prompt and various ways to control transcription style. This prompt feature of Whisper has been used in prior work \cite{application3} for a variety of tasks in a zero-shot manner such as audio-visual speech recognition, wherein the visual context is converted to a list of bias words using CLIP \cite{clip}, which is then integrated to prompt, improving transcription accuracy over using audio alone. 

However, Whisper was not trained to follow any particular instruction\footref{myfootnote}. The model has been trained to use the previous speech segment's transcription as prompt for transcribing the current speech segment, similar to how a base GPT (Generative Pre-trained Transformers) \cite{gpt2} model generates the next series of tokens given previously entered text. Hence, Whisper's prompt operates similar to a base GPT model\footref{myfootnote}, such as GPT-2, in that although Whisper is not explicitly trained on instructions, it can still follow an instruction given in the prompt in a zero-shot manner. 

Previous works have shown that fine-tuning base GPT models with instructions significantly increases their ability to follow instructions and improves their zero-shot performance across various datasets and tasks \cite{humanfeedback}. In this paper, we leverage this concept of instruction-tuning to fine-tune Whisper for a single instruction of contextual biasing, particularly for rare words. Additionally, we investigate whether fine-tuning Whisper for biasing instruction on a single English dataset leads to generalization across other English datasets. To this end, we present a novel data-efficient supervised learning method to improve the performance of Whisper for contextual biasing through prompting. We propose a specific prompt selection strategy aimed at more robust training, and a weighting scheme for the Cross Entropy loss for maximizing learning for the given task. We name our fine-tuned model \textit{Bias-Whisper} or \textit{B-Whisper}.

 Giving credence to our hypothesis, our experiments suggest that despite using only $670$ hours of Common Voice English set \cite{cv} for fine-tuning, our model generalizes for biasing instruction to $11$ open-source datasets such as TED-LIUM \cite{tedlium}, SLURP \cite{slurp}, Vox Populi \cite{vox}, etc., outperforming the baseline by a significant margin in terms of WER for rare words. Surprisingly, for contextual biasing, the model shows increased performance even on words not seen during fine-tuning. Moreover, despite being fine-tuned only on an English dataset, our model performs well on unseen languages like French, Spanish, Italian, and German for biasing instruction---further testifying to the zero-shot capabilities of our model. Notably, existing research on contextual biasing has predominantly focused on English, leaving a gap in studies addressing other languages. Our study reduces this gap by showcasing a method that generalizes to unseen languages although trained on an English dataset.

\section{Related Work}

In this section, we discuss recent research diving into the prompting capabilities of Whisper. 
This work \cite{application3} is among the first to demonstrate the zero-shot capabilities of prompting in Whisper. Specifically, for various tasks such as audio-visual speech recognition, code-switched speech recognition, and speech translation, this paper presented a way contextual knowledge can be integrated into the Decoder through prompt, which significantly enhances Whisper's transcription accuracy. Another paper \cite{speakerprompt} proposed a prompt-tuning methodology that adds prompts to both the Encoder and Decoder parts of the model to transcribe the target speaker's speech from overlapped multi-talker audios.

The closest work to our approach is \cite{zeroshotdomain}, which introduces a method for creating domain-sensitive Whisper by fine-tuning it on textual prompts that describe the audio context and genre. However, the paper's definition of `prompt' is somewhat vague and domain-dependent, potentially resulting in subjective variations in model performance.

\begin{figure*}
  \centering
  \includegraphics[scale=0.95]{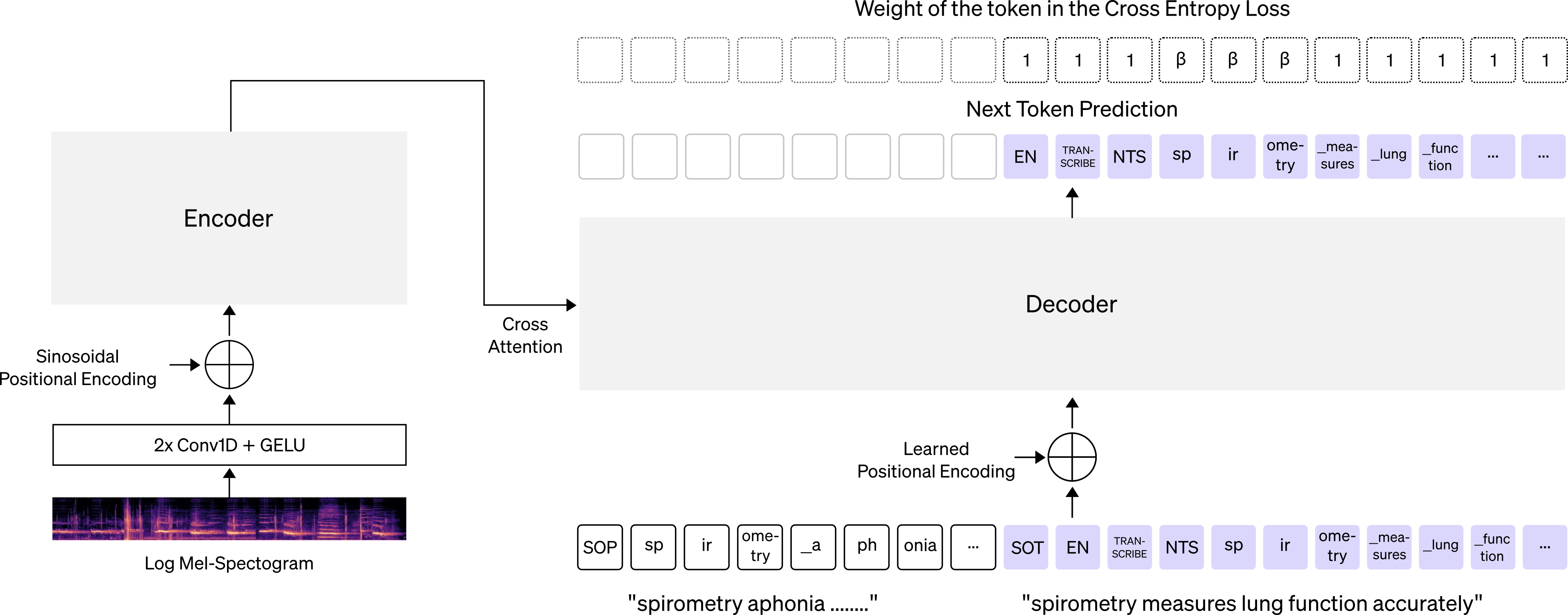} 
  \label{fig:a}
  \caption{\textbf{Overview of B-Whisper.} This figure depicts the train-time inputs and outputs for B-Whisper with an example. In the bottom-most sequence of squares, colored squares contain task-specifiers and transcript tokens, whereas colorless squares contain prompt tokens. In this case, the word ``spirometry'' is the true-bias word for the reference transcript ``spirometry measures lung function accurately''. As such, the tokens of ``spirometry'' are given weight $\beta > 1$ and the rest of the tokens are given weight $1$ in Cross Entropy Loss, as shown in the top-most sequence of squares. }\label{fig:AB}
\end{figure*}

\section{Background}

In this section, we briefly discuss Whisper and its architectural details \cite{whisper}. Whisper is a family of sequence-to-sequence models that follow the encoder-decoder Transformer architecture \cite{attention}. Instead of a gold standard human-validated dataset, Whisper has been trained in a weakly-supervised fashion on $680,000$ hours of web-scaled speech dataset. One of the distinctive features of the Whisper model is its ability to handle a variety of speech-related tasks, such as transcription, translation, language identification, and voice activity detection (VAD).

Given an input speech signal, its log mel-spectogram is first calculated, denoted as $A_{T \times M}$, where $M$ is the number of mel-bins and $T$ is the total number of frames. These features are then passed through the encoder $E$, yielding speech representations $h$ for the given speech signal.

\begin{equation}
    h = E(A)
\end{equation}

The speech representations $h$, along with the sequence of prompt tokens $c=\{c_{1}, c_{2}, , ...,  c_{C}\}$ are then passed to the decoder $D$.  The decoder generates probability distribution $p_i$ for the next token $y_i$, conditioned on the previously decoded outputs $y_{<i}$, speech representations, and sequence of prompt tokens:

\begin{equation} \label{eq:1}
    p_i = P(y_i|y_{<i},h,c)
\end{equation}

For more details regarding the multitasking format through prompt tokens or Whisper's training details, see \cite{whisper}.

\section{Methodology}

In this section, we introduce a supervised learning scheme to fine-tune Whisper for a single instruction for contextual biasing, specifically for rare words.
We now outline our approach for creating the training prompt for a particular utterance and provide details regarding the loss function used for supervised learning.

\textbf{Prompt Selection Strategy:} Prior to training, for each speech utterance, its reference transcript and hypothesis transcript generated using base Whisper model are aligned to find out incorrectly transcribed rare words in the reference transcript. A word is considered a rare word if it falls outside the most common words accounting for $90\%$ of word occurrences in the train set. Out of these incorrectly transcribed rare words, one word is randomly selected to be the true-bias word for that utterance. By including incorrectly transcribed rare words as true-bias words in the biasing list, we can focus the fine-tuning on resolving base Whisper's mistakes, hence allowing the model to learn to better utilize the biasing list to correct its mistakes. Moreover, since we are unaware of the frequency distribution of words in Whisper’s original training data, we believe selecting true-bias words based on incorrect transcription and rarity rather than rarity alone would lead to better results \cite{tcpgen}.

We maintain a global list $V$ of such incorrectly transcribed words that fall outside the most common words accounting for $90\%$ of word occurrences in the training dataset.  For each utterance, we randomly select $L$ false-bias words from this list $V$, which do not occur in the given reference transcript. Here $L$ is randomly chosen from $[25, 150]$ so as to ensure varying biasing list size in training. To enhance the resilience of the model in cases where the speech utterance does not contain any word present in the biasing list, we randomly drop the true-bias words from the list with probability $P_{neg}$, hence keeping the biasing list full of false-bias words only in such a scenario. In addition to this, to ensure the model performs well even when no biasing list is given, we completely drop the biasing list with probability $P_{empty}$. We combine the selected false-bias and true-bias words in a random order to create the biasing list. Then, we concatenate all the words in the biasing list using ``space'' characters to form the textual prompt for the given speech utterance.

\textbf{Weighted Cross Entropy (CE) Loss:} The number of rare words in an utterance is generally in minority compared to the total number of words present in it. Hence, to prioritize learning for such words, we introduce weight in the CE Loss. We increase the weight of the corresponding tokens for the true-bias words to $\beta > 1$ and keep the weight corresponding to the other tokens to $1$. The modified loss becomes:

\begin{equation}
    L = \sum_{i=1}^{S} w_i H(q_i, p_i)
\end{equation}

\noindent where $H$ is the cross entropy function, $p_i$ is as defined in (\ref{eq:1}), $q_i$ is the one-hot ground truth vector, $w_i$ is $\beta >1$ if $y_i$ is a token of true-bias word, else it is $1$, and $S$ is the total number of tokens in the transcript. This choice of prompt selection strategy and loss makes the model learn Whisper's mistakes effectively during fine-tuning. Figure \ref{fig:AB} shows our approach.

\section{Experimental Settings}

\textbf{Train Dataset:} We used the Common Voice \cite{cv} English $17.02$ dataset for training, which consists of $2615$ hours of  labeled audios from over $90k$ global speakers, providing a broad range of accents and speakers---which can help for effective model generalization. In our experiments, we used a subset of approximately $670$ hours from the Common Voice training split as our train set, and the official val split for validation. We restricted to a smaller train set to prevent excessive bias towards the training distribution, thereby maintaining Whisper’s generalized performance. Prior to training, we also normalized all the reference transcripts with the English normalizer function available in the official Whisper repository. \footnote{https://github.com/openai/whisper/}

\textbf{Test Datasets:} In addition to testing on the Common Voice test set and the entire Artie Bias \cite{artie} which is a subset of the Common Voice English dataset, we were also interested in comprehensively evaluating our model on out-of-domain datasets. Hence, we chose a set of $9$ open-source out-of-domain English datasets for testing: official evaluation set of Chime6 \cite{chime}, CORAAL:VLD v. 2021.07 component of CORAAL \cite{coraal}, test set of SLURP \cite{slurp}, test set of TED-LIUM \cite{tedlium}, English test set of VoxPopuli \cite{vox}, test-clean and test-other splits of LibriSpeech \cite{Libri}, English test set of FLEURS \cite{fleurs}, and test set of ``Medical Speech, Transcription, Intent'' \cite{medical}. In addition to these, we also evaluated on the French (fr), Spanish (es), Italian (it), and German (de) test splits of Multilingual LibriSpeech (MLS) \cite{MLS}. All the evaluations on these out-of-domain datasets were done in a zero-shot setting, without using the training set of the specific dataset during model fine-tuning, to evaluate generalization. Additionally, none of these sets were used in the training of Whisper. We normalized the reference transcripts for all 11 datasets as recommended in \cite{whisper}.

\textbf{Test-Time Biasing List Creation:} We test on different biasing list conditions to ensure the model's robustness across various scenarios. We adopt two scenarios for the creation of the biasing list.

In Scenario-1, we build the biasing list along the lines of \cite{tcpgen, tang2024improving, trie}, wherein for each utterance, for true-bias words, we extract words in the reference transcript that fall outside the most common words accounting for $90\%$ of word occurrences in the train set. We sample false-bias words from a global list of rare words---words that are not present in the most common words accounting for $90\%$ of word occurrences in the corresponding train set.

In Scenario-2, we aim to emulate the instances where none of the words in the biasing list are spoken in the given audio. To this end, following along the lines of \cite{text-injection}, we create the biasing list which only contains false-bias words sampled randomly from the global list of rare words.

\textbf{Model Details:} We fine-tuned our model on top of the pre-trained Whisper Large architecture initialized with the official `openai/whisper-large' checkpoint. The learning rate was set to $10^{-7}$ with Adam optimizer and a linear rate decay. The dropout rate was set to $10\%$. We trained our model for $1$ epoch. We extended the positional embedding enough to accommodate $756$ tokens. The value of $\beta$ for weighted Cross Entropy Loss was set to $1.1$. For all the experiments, feature extraction, pre-processing, and tokenization steps were the same as mentioned in \cite{whisper}. For biasing list selection during training, we chose $P_{neg}=0.3$ and $P_{empty}=0.2$. For transcription generation for both B-Whisper and Whisper, we kept beam size $1$. 

\textbf{Evaluation Metrics:} In line with previous works \cite{tcpgen, trie, sathyendra2022contextual, shakeel2024contextualized, pointer} we use the following four evaluation metrics - WER: word error rate for all the words, U-WER: unbiased word error rate, or word error rate for words not part of the biasing list, R-WER: rare word error rate or word error rate for words present in the biasing list, and OOV-WER: word error rate for OOV (out-of-vocabulary) words present in the biasing list. Here, OOV words refer to the words that are not present in our training word vocabulary, whose size is around $230K$. However, these words might be present in the pre-training dataset for Whisper. Lower is better for all the mentioned evaluation metrics. We report all the mentioned metrics in \%.

\begin{table*}[t]
\vskip 0.15in
\small
\begin{center}
\begin{tabular}{l|l|cc|ccccccccc|c}
\toprule
  & & \rotatebox{90}{\CellWithForceBreak{Common\\ Voice}} &  \rotatebox{90}{Artie Bias}  & \ \rotatebox{90}{Chime6} 
 & \rotatebox{90}{CORAAL}  & \rotatebox{90}{FLEURS}  & \rotatebox{90}{LS-Clean}  & \rotatebox{90}{LS-Other}  & \rotatebox{90}{Medical}  & \rotatebox{90}{SLURP}  & \rotatebox{90}{TED-LIUM}  & \rotatebox{90}{VoxPopuli} &\rotatebox{90}{Average}  \\
\midrule

\multirow{3}{*}{WER} & Whisper  & 11.0 & 6.7 & 25.3 & \textbf{19.7} & 6.3 & 2.6 & 5.4 & 8.4 & 15.8 & \textbf{4.6} & 7.1 & 10.3 \\
& Whisper + P  & 10.1 & 6.3 & 32.5 & 28.3 & 6.0 & 2.2 & 4.6 & 8.5 & 16.3 & 8.3 & 9.6 & 12.1 \\
& B-Whisper & 10.0 & 6.4 & 23.9 & 20.9 & 6.6 & 2.7 & 5.5 & 8.3 & 16.2 & 4.8 & 7.8 & 10.3 \\
& B-Whisper + P  & \textbf{7.0} & \textbf{4.6} & \textbf{22.7} & 20.4 & \textbf{5.2} & \textbf{1.5} & \textbf{3.4} & \textbf{6.4} & \textbf{14.9} & 4.7 & \textbf{6.9} & \textbf{8.9}\\

\midrule

\multirow{3}{*}{U-WER} & Whisper & 8.6 & 5.3 & 24.5 & \textbf{18.4} & 5.4 & 1.7 & 3.6 & 7.3 & \textbf{14.4} & \textbf{4.4} & \textbf{6.8} & 9.1 \\
& Whisper + P & 9.3 & 5.9 & 32.5 & 27.5 & 5.5 & 1.8  & 3.8 & 7.9  & 15.8 & 8.1 & 9.4 & 11.6 \\
& B-Whisper & 7.5 & 5.0 & 22.7 & 19.6 & 5.4 & 1.6 & 3.7 & 7.2 & 14.7 & 4.6 & 7.4 & 9.0 \\
& B-Whisper + P & \textbf{6.8} & \textbf{4.7} & \textbf{22.7} & 19.8 & \textbf{5.2} & \textbf{1.5} & \textbf{3.3} & \textbf{6.1} & 14.5 & 4.7 & 6.9 & \textbf{8.7}\\

\midrule

\multirow{3}{*}{R-WER} & Whisper & 35.7 &  22.4 & 32.1 & 44.1 & 16.8 & 10.9 & 21.5 & 21.2 & 34.3 & 10.1 & 12.0 & 23.7 \\
& Whisper + P & 18.6 & 10.7 & 32.4 & 44.9 & 11.3 & 6.4 & 11.6 & 15.2 & 22.9 & 11.8 & 12.6 & 18.0 \\
& B-Whisper & 35.8 & 22.0 & 34.7 & 47.3 & 21.2 & 11.5 & 21.4 & 20.3 & 35.8 & 9.8 & 14.2 & 24.9 \\
& B-Whisper + P & \textbf{8.7} & \textbf{4.0} & \textbf{23.7} & \textbf{32.9} & \textbf{5.5} & \textbf{2.2} & \textbf{5.0} & \textbf{10.5} & \textbf{19.5} & \textbf{5.2} & \textbf{5.5} & \textbf{11.2}  \\

\midrule
R-WERR &  & 53.2 & 62.6 & 26.8 & 26.7 & 51.3 & 65.6 & 56.9 & 30.9 & 14.9 & 55.9 & 56.3 & \color{darkgreen}45.6 \\

\midrule

\multirow{3}{*}{OOV-WER} & Whisper & 75.7 & 62.1 & 62.2 & 73.5 & 58.0 & 51.5 & 64.5 & 72.1 & 68.0 & 23.0 & 49.0 & 60.0 \\

& Whisper + P & 33.2 & 24.3 & 46.8 & 64.9 & 38.3 & 24.4 & 30.2 & 54.4 & 37.8 & 23.0 & 30.9 & 37.1\\
& B-Whisper & 71.9 & 56.8 & 58.5 & 76.4 & 63.0 & 52.0 & 65.2 & 71.7 & 68.8 & 29.2 & 50.5 & 60.4 \\
& B-Whisper + P & \textbf{8.7} & \textbf{5.4} & \textbf{27.6} & \textbf{38.2} & \textbf{14.8} & \textbf{7.2} & \textbf{10.2} & \textbf{29.3} & \textbf{17.3} & \textbf{6.1} & \textbf{11.4} & \textbf{16.0}\\

\midrule
OOV-WERR &  & 73.8 & 77.8 & 41.0 & 41.1 & 61.4 & 70.5 & 66.2 & 46.1 & 54.2 & 73.5 & 63.1 & \color{darkgreen}60.8 \\

\bottomrule

\end{tabular}
\caption{Overview of various WER metrics of Whisper without biasing list, Whisper with biasing list ($N = 70$), B-Whisper without biasing list, and B-Whisper with biasing list ($N = 70$). Here, R-WERR denotes the relative percentage reduction in R-WER of B-Whisper+P compared to the Whisper+P. The same applies to OOV-WERR.}
\label{robustness_table}
\end{center}
\vspace{-1em}
\end{table*}

\section{Results}

\textbf{Effect of Contextual Biasing: } We have comprehensively compared the performance of Whisper and B-Whisper across $11$ open-source English datasets, out of which for $9$ datasets in a zero-shot setting, the results of which are given in Table \ref{robustness_table}. Here ``Whisper'' and ``B-Whisper'' denote the use of these models without providing any biasing list, whereas ``Whisper+P'' and ``B-Whisper+P'' denote the use of biasing list via prompting (biasing list size $N=70$). We follow this convention for the rest of the paper. It should be noted that even when we did not pass any biasing list to Whisper and B-Whisper, we used the corresponding biasing list used in Whisper+P and B-Whisper+P in order to calculate their U-WER, R-WER, and OOV-WER values. The biasing list used for Table \ref{robustness_table} was created as mentioned in Scenario-1. 

\begin{table*}[t]
\vskip 0.15in
\small
\begin{center}
\begin{tabular}{|>{\centering\arraybackslash}m{2.5cm}|>{\centering\arraybackslash}m{2.5cm}|>{\centering\arraybackslash}m{2.5cm}|>{\centering\arraybackslash}m{2.5cm}|>{\centering\arraybackslash}m{2.5cm}|>{\centering\arraybackslash}m{2.5cm}|}
    \hline
    \textbf{Transcript} & \textbf{Whisper} & \textbf{B-Whisper} & \textbf{Whisper+P} & \textbf{B-Whisper+P} & \textbf{Biasing List} \\

    \hline
    ... foreign rule to the \textbf{phanariote} period & ... foreign rule to the \textbf{\textcolor{darkred}{fanaret}} period & ...  foreign rule to the \textbf{\textcolor{darkred}{fanaret}} period & ...  foreign rule to the \textbf{\textcolor{darkred}{phanaret}} period & ... foreign rule to the \textbf{\textcolor{darkgreen}{phanariote}} period & [..,mcphillips, \textbf{phanariote}, lukyamuzi,..] \\ 
    \hline
     i feel pain in my ears with \textbf{tinnitus} & i feel pain in my ears with \textbf{\textcolor{darkred}{cheetahs}} & i feel pain in my ears with \textbf{\textcolor{darkred}{cheetahs}} &  epilpian in my ears with \textbf{\textcolor{darkred}{cheetahs}} & i feel pain in my ears with \textbf{\textcolor{darkgreen}{tinnitus}} &  [..,kimbolton, \textbf{tinnitus}, polygynandy,..] \\
    \hline
    
\end{tabular}
\caption{Example transcripts of various models on samples from Common Voice (first) and Medical test set (second).}
\label{tab:dummy_data}
\end{center}
\vspace{-1em}
\end{table*}

\begin{table}[t]
\vskip 0.15in
\small
\begin{center}
\begin{tabular}{l|l|cccc}
\toprule
  & & fr & de & es & it \\
\midrule

\multirow{4}{*}{WER} 
& Whisper         & 8.3 & 6.3 & 4.4 & 12.9 \\
& Whisper + P     & 10.8 & 15.2 & 5.9 & 13.6 \\
& B-Whisper       & 7.4 & 7.6 & 5.1 & 14.8 \\
& B-Whisper + P   & \textbf{5.5} & \textbf{5.5} & \textbf{3.7} & \textbf{11.6} \\
\midrule

\multirow{4}{*}{U-WER} 
& Whisper         & 7.6 & \textbf{6.0} & \textbf{4.0} & \textbf{13.6} \\
& Whisper + P     & 10.4 & 15.0 & 6.3 & 15.4 \\
& B-Whisper       & 6.3 & 7.2 & 4.7 & 15.7 \\
& B-Whisper + P   & \textbf{5.9} & 7.1 & 4.2 & 14.3 \\
\midrule

\multirow{4}{*}{\CellWithForceBreak{OOV- \\ WER}} 
& Whisper         & 17.3 & 15.6 & 10.3 & 19.6 \\
& Whisper + P     & 21.0 & 27.7 & 8.7 & 12.7 \\
& B-Whisper       & 18.8 & 17.2 & 10.9 & 20.2 \\
& B-Whisper + P   & \textbf{6.7} & \textbf{4.2} & \textbf{3.4} & \textbf{5.2} \\
\bottomrule

\end{tabular}
\caption{Results for Whisper and B-Whisper (with and without biasing list with $N=70$) on various test sets of MLS.}
\label{multi_table}
\end{center}
\vspace{-1em}
\end{table}

 \begin{figure*}[t]
  \centering
  \includegraphics[width=\linewidth]{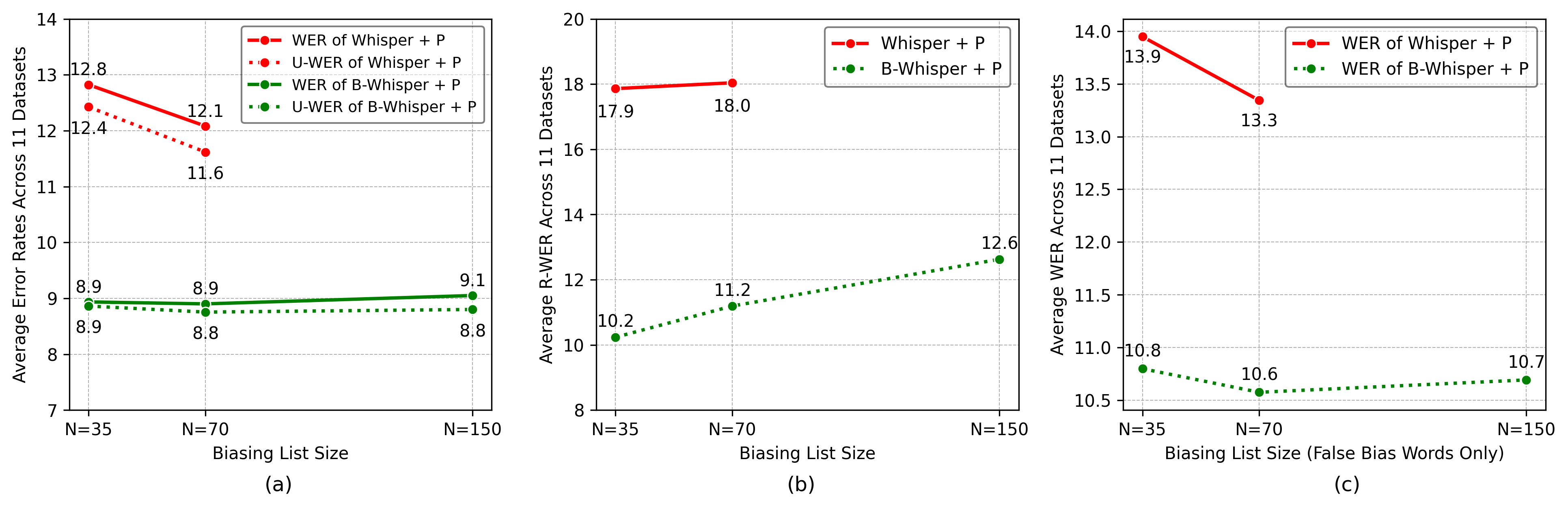}
  \caption{Average values across 11 datasets of (a) WER and U-WER, (b) R-WER, (c) WER, for biasing list sizes $35$, $70$ and $150$ for Whisper + P and B-Whisper + P. Here, in case of (c), the biasing list contains only false-bias words.}
  \label{fig:prompt_lengths}
\end{figure*}

Comparing the performance of Whisper+P with Whisper, Whisper+P sees a clear reduction in average R-WER from $23.7\%$ to $18.0\%$, and average OOV-WER from $60\%$ to $37.1\%$. This demonstrates the usefulness of using prompt `out-of-the-box' for contextual biasing in Whisper. However, U-WER of Whisper+P increases in comparison with Whisper across all the datasets, resulting in an increased WER across $6$ out of $11$ datasets in  comparison to Whisper. This increase in U-WER and overall WER can be attributed to the fact that Whisper's prompt expects the transcription of the previous speech segment rather than a list of biasing words. In other words, the prompt for Whisper is misaligned for the instruction of contextual biasing. 

Compared to Whisper+P, R-WER reduces on average by $45.6\%$ for B-Whisper+P. Although we only used Common Voice to train our model, the improvement is consistent across all the datasets as evident from the R-WERR values, thus proving the effectiveness of our approach in a zero-shot setting. Even in the case of OOV-WER, where the words to be biased are completely absent from the fine-tuning train set, we see that B-Whisper+P has achieved the best performance across all the datasets, with an average improvement of $60.8\%$ over Whisper +P. This shows the strong generalization ability of B-Whisper+P. An interesting observation here is that the relative improvement in R-WER and OOV-WER is higher in datasets where Whisper already performs relatively well, such as Artie Bias and VoxPopuli. Examples of predictions illustrating the effectiveness of B-Whisper+P are given in Table \ref{tab:dummy_data}.

Moreover, B-Whisper+P largely fixes the poor performance of Whisper+P when it comes to U-WER and WER, achieving the best results in U-WER for $7$ and WER for $9$ out of $11$ datasets. When used without prompting, B-Whisper achieves near identical WER as that of Whisper in most datasets, with slight deviations possibly due to distribution shift during fine-tuning. Additionally, the hyper-parameter $P_{empty}$ can also influence these results. Overall, this shows that B-Whisper has almost retained its original behaviour in conditions where no biasing list is provided, in spite of fine-tuning.

\textbf{Impact of Biasing List Size: } To analyze the effect of biasing list size on contextual biasing capabilities, we have plotted Figure \ref{fig:prompt_lengths} (a) and (b), which shows the average values of WER, U-WER, and R-WER across $11$ datasets of Whisper+P and B-Whisper+P for varying biasing list size. The biasing list size is limited to $70$ for Whisper since by default it allows a maximum of $224$ tokens as input for prompt, which approximates to around $70$ words.  We can see that both models suffer from degradation in R-WER as biasing list size increases, as the number of false-bias words has increased. This change is more pronounced in B-Whisper+P. However, the values of R-WER for B-Whisper+P are around $7\%$ points lower than Whisper+P, on account of the supervised fine-tuning. Surprisingly, in the case of U-WER for Whisper+P, there is an improvement when going from $N=35$ to $N=70$. On the other hand, B-Whisper shows a near-consistent U-WER with an increase in $N$. This could be attributed to our training regime wherein the biasing list size is randomly chosen for each training sample, making it more robust to changes in biasing list size. 

To gauge the effect of over-biasing with different biasing list sizes, Figure \ref{fig:prompt_lengths} (c) shows the average WER across $11$ datasets for both Whisper+P and B-Whisper+P. The biasing lists, as mentioned in Scenario-2, contain only false-bias words. For both the models, we see a decrease in WER when going from $N=35$ to $N=70$. 

However, the WER for B-Whisper+P is around $3\%$ points lower, indicating its effectiveness in handling purely false bias words in a biasing list.

Moreover, it is interesting to note that when going from $N=70$ to $N=150$, there is a minimal increase in WER. This shows that B-Whisper is largely resistant to over-biasing with different biasing list sizes.

\textbf{Evaluation on Unseen Languages: } In order to evaluate the biasing capabilities of our model on languages unseen during fine-tuning, we performed evaluation on four European languages, the results of which have been summarized in Table \ref{multi_table}. Although contextual biasing via prompting is effective in Whisper for English, it proves ineffective for certain non-English languages such as French and German, hence degrading both R-WER and U-WER. This might be due to the Whisper model's pre-training dataset, which primarily consists of English data, along with the misalignment in prompt definition as mentioned earlier. To our surprise, in contrast to Whisper+P, B-Whisper+P has been able to successfully translate its biasing capabilities from English to these languages, achieving better OOV-WER. Consequently, it has seen the best WER in all four languages. On the other hand, it has not seen a similar improvement in U-WER, where it lags slightly behind Whisper in $3$ out of $4$ languages. 

Overall, our results show that B-Whisper performs well not only on its fine-tuning test set but also adeptly integrates broad knowledge acquired during pre-training of Whisper with specific skills acquired through our fine-tuning procedure, thereby generalizing on unseen languages for a new instruction. Although we keep a maximum biasing list size of $150$ in our experiments, this can be extended by a retriever mechanism to filter out relevant words, similar to what is done in \cite{saket}.

\section{Conclusion}

This paper explores the impact of fine-tuning Whisper for contextual biasing instruction, specifically for rare words in zero-shot settings. Our main finding is that despite using a small set of $670$ hours English dataset for fine-tuning, our model B-Whisper outperforms Whisper by a large margin on $11$ open-source English datasets and also on languages unseen during fine-tuning process. Due to its generalized performance despite using a relatively small amount of training data, this approach can be particularly valuable for ASR practitioners and researchers struggling with low accuracy on domain-specific terms in Whisper—especially those with limited access to extensive computational resources or large industrial-scale labeled datasets. Another exciting future work could be extending our approach by fine-tuning Whisper on multiple instructions useful for tasks such as audio-visual speech recognition, code-switching speech recognition, etc.

\section{Acknowledgments}

We greatly appreciate Yoginkumar Patel for his helpful comments and support throughout this work.

\label{sec:print}

\bibliographystyle{IEEEbib}
\bibliography{mybib}
\end{document}